\documentclass[conference,letterpaper]{IEEEtran}
\usepackage{graphicx,psfrag,epsfig,epsf,latexsym,hhline,amssymb,multirow}
\usepackage[utf8]{inputenc}
\usepackage[T1]{fontenc}
\usepackage[cmex10]{amsmath}
\usepackage{pst-plot}
\usepackage{pstricks-add}
\usepackage{pifont}
\usepackage{graphicx}
\usepackage{amsthm}
\usepackage{algorithm}
\usepackage{algorithmic}
\usepackage[noadjust]{cite}
\usepackage{array}
\usepackage{blindtext}
\usepackage{etoolbox}
\usepackage{comment}
\usepackage{epstopdf}
\usepackage{hyperref}

\newcolumntype{M}[1]{>{\centering\arraybackslash}m{#1}}
\newcolumntype{P}[1]{>{\centering\arraybackslash}p{#1}}

\newcommand{\iid}{i.\@i.\@d.\ }

\theoremstyle{definition}
\theoremstyle{definition}
\theoremstyle{definition}
\theoremstyle{definition}

\IEEEoverridecommandlockouts
\begin{document}
\title{Generalized Spatially Coupled Parallel Concatenated Convolutional Codes With Partial Repetition}
\author{
\IEEEauthorblockN{Min Qiu, Xiaowei Wu, Jinhong Yuan}
\IEEEauthorblockA{University of New South Wales\\
Sydney, Australia\\
\{min.qiu, xiaowei.wu, j.yuan\}@unsw.edu.au}   %<------ Line breaks in the current column
\and
\IEEEauthorblockN{Alexandre Graell i Amat}
\IEEEauthorblockA{Chalmers University of Technology\\
Gothenburg, Sweden\\
alexandre.graell@chalmers.se}   %<------ Line breaks in the current column
%\and
%\IEEEauthorblockN{Xiaowei Wu and Jinhong Yuan}
%\IEEEauthorblockA{University of New South Wales\\
%Sydney, Australia\\
%\{xiaowei.wu, j.yuan\}@unsw.edu.au}  %<------- Extra vertical space
%\and
%\IEEEauthorblockN{Jinhong Yuan}
%\IEEEauthorblockA{University of New South Wales\\
%Sydney, Australia\\
%j.yuan@unsw.edu.au}
\thanks{The work of M. Qiu, X. Wu and J. Yuan was supported in part by the Australian Research Council Discovery Project under Grant DP 190101363 and in part by the Linkage Project under Grant LP 170101196. A. Graell i Amat was supported by the Swedish Research Council under Grant 2020-03687.}

%\thanks{This work was presented in part at xxx.
%
%
%The authors are with xxx.
%%
%%Y.-C. Huang is with the Department of Communication Engineering, National Taipei University, New Taipei City 23741, Taiwan (e-mail: ychuang@mail.ntpu.edu.tw).
%}%
}

\maketitle

\begin{abstract}
We introduce generalized spatially coupled parallel concatenated codes (GSC-PCCs), a class of spatially coupled turbo-like codes obtained by coupling parallel concatenated codes (PCCs) with a fraction of information bits repeated before the PCC encoding. GSC-PCCs can be seen as a generalization of the original spatially coupled parallel concatenated convolutional codes (SC-PCCs) proposed by Moloudi \emph{et al.} \cite{8002601}. To characterize the asymptotic performance of GSC-PCCs, we derive the corresponding density evolution equations and compute their decoding thresholds. We show that the proposed codes have some nice properties such as threshold saturation and that their decoding thresholds improve with the repetition factor $q$. Most notably, our analysis suggests that the proposed codes asymptotically approach the capacity as $q$ tends to infinity with any given constituent convolutional code.
\end{abstract}

%\begin{IEEEkeywords}
%Spatial coupling, turbo codes, density evolution.
%\end{IEEEkeywords}

\section{Introduction}
Turbo codes \cite{397441,Vucetic:2000:TCP:352869} and low-density parity-check (LDPC) codes \cite{Gallager63low-densityparity-check} are two important classes of codes that have been widely used in many communication systems. These codes are capable of achieving near-Shannon-limit performance as the block length grows large. Spatial coupling \cite{782171,5571910} brings further performance improvement to these codes. Particularly, it was analytically proven in \cite{5695130} that spatially coupled LDPC codes with suboptimal belief propagation (BP) decoding achieve the optimal maximum-a-posteriori (MAP) decoding threshold, a phenomenon known as threshold saturation \cite{5695130}. Since then, the concept of spatial coupling yields an attractive way for constructing capacity-approaching channel codes from various component codes.

In this work, we focus on turbo-like codes whose factor graphs \cite{910572} have convolutional code trellis constraints. Turbo-like codes include parallel concatenated codes (PCCs) \cite{397441}, serially concatenated codes (SCCs) \cite{669119} and braided convolutional codes (BCCs) \cite{5361461}. Spatially coupled turbo-like codes were introduced in \cite{8002601} by applying spatial coupling to turbo-like codes. It was proven in \cite{8002601} that threshold saturation also occurs for this class of codes. Further investigations on the tradeoff between error floor and waterfall performance of spatially coupled turbo-like codes as well as the effects of coupling memory and component code block length on decoding performance were conducted in \cite{8631116} and \cite{mahdavi2020effect}, respectively. Despite the capacity approaching performance of SC-SCCs and SC-BCCs, the thresholds of SC-PCCs (when punctured) are (strictly) bounded away from the capacity \cite{8002601}. Aiming to improve the performance of PCCs, partially-information coupled turbo codes (PIC-TCs), where spatial coupling is applied on the turbo code level instead of the convolutional code level as in \cite{8002601}, were proposed in \cite{8368318}. The main idea is that adjacent code blocks share a fraction of information bits such that these bits are protected by two component turbo codewords. A similar idea based on partially coupling parity bits was proposed in \cite{PIC2020}. One benefit of such construction is that the technique of partial coupling can be applied to any component code without changing its encoding and decoding architecture. It was shown that partially coupled turbo codes \cite{8989359,PIC2020,9174156} can attain significant threshold improvements over uncoupled turbo codes, although threshold saturation was neither observed nor proved. Since PCCs are still the standard channel codes in current wireless mobile communication systems, we are interested in designing new coupled codes with PCCs as component codes that are compatible with the current standard.

In this paper, we introduce generalized SC-PCCs (GSC-PCCs), which are constructed by applying spatial coupling on a component PCC, where a fraction of information bits are repeated. The proposed codes not only can be seen as a generalization of the original SC-PCCs \cite{8002601}, but also exhibit a similar structure to that of PIC-TCs \cite{PIC2020} as the repeated bits are protected by the component PCC codewords at several time instants. We derive the exact density evolution equations for both uncoupled and coupled ensembles and optimize the fraction of information repetition to obtain the largest decoding thresholds on the binary erasure channel (BEC). We analytically show that threshold saturation occurs for GSC-PCCs by using potential functions \cite{8002601}. Numerical calculations reveal that the decoding threshold of GSC-PCCs improves with increasing repetition factor $q$ and is better than that of SC-PCCs \cite{8002601} and PIC-TCs \cite{PIC2020} even with $q=2$. Both our analysis and numerical results suggest that GSC-PCCs approach the BEC capacity asymptotically as $q$ tends to infinity with any given constituent convolutional code regardless of its code rate and number of states.

%Since GSC-PCCs contain the features of both SC-PCCs \cite{8002601} and PIC-TCs \cite{PIC2020}, the proposed codes can achieve even better decoding performance than the aforementioned two codes. Most importantly,

%\subsection{Notation}
%Scalars and vectors are written in lightface and boldface letters, respectively, e.g., $x$ and $\mathbf{x}$.

\begin{figure}[t!]
	\centering
\includegraphics[width=2.5in,clip,keepaspectratio]{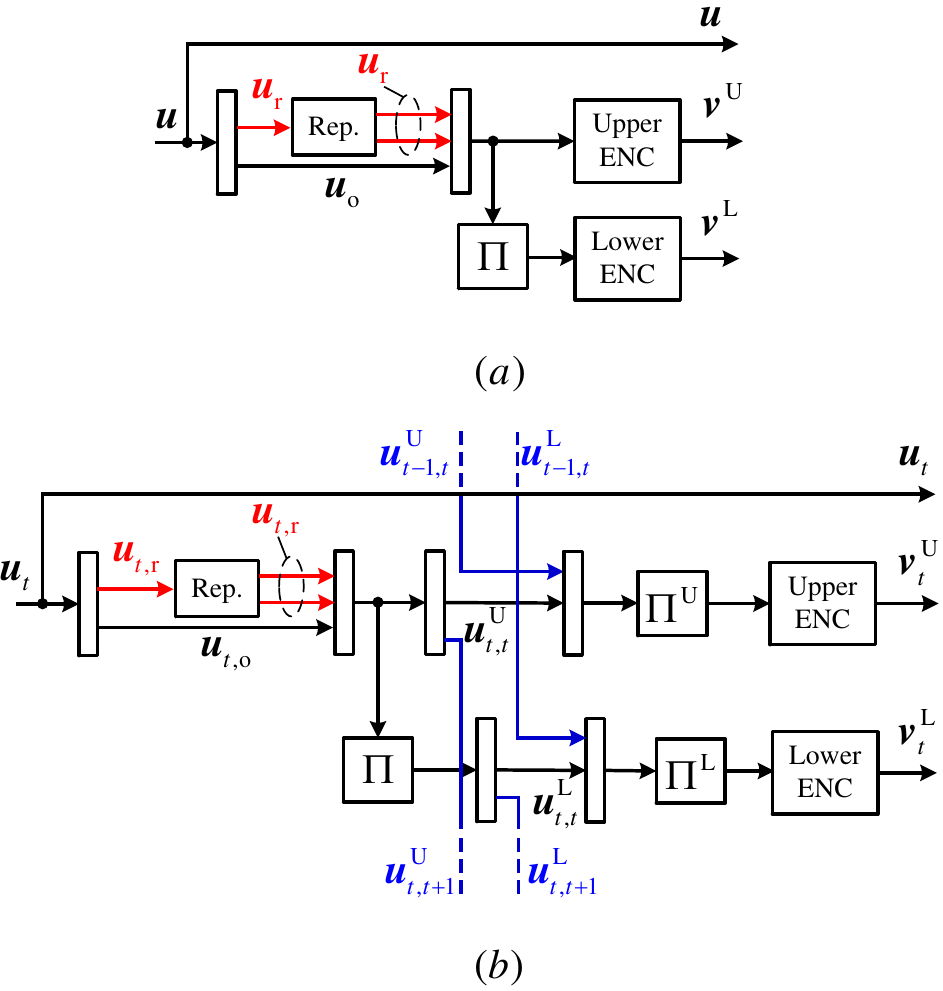}
\vspace{-3mm}
\caption{Block diagram of the encoder of (a) an uncoupled PCC with partial information repetition, and (b) a GSC-PCC with $m=1$ at time $t$.}
\label{fig:GSC_PCC_ENC}
\vspace{-4mm}
\end{figure}

\section{Generalized Spatially Coupled Parallel Concatenated Codes}
In this section, we introduce the construction of GSC-PCCs. Before proceeding, we briefly introduce the uncoupled PCCs with partial information repetition that are used as the component codes of GSC-PCCs. The uncoupled codes are similar to the dual-repeat-punctured turbo codes in \cite{5308225}, except that in our case only a fraction of the information bits are repeated. The encoder of the uncoupled PCC with partial repetition is depicted in Fig. \ref{fig:GSC_PCC_ENC}(a). A length-$K$ information sequence $\boldsymbol{u}$ is divided into two sequences, $\boldsymbol{u}_{\text{r}}$ and $\boldsymbol{u}_{\text{o}}$. Then, sequence $\boldsymbol{u}_{\text{r}}$ is repeated $q$ times. After repetition, the length-$K'$ information sequence $[\boldsymbol{u}_{\text{r}},\ldots,\boldsymbol{u}_{\text{r}},\boldsymbol{u}_{\text{o}}]$ and its reordered copy $\Pi([\boldsymbol{u}_{\text{r}},\ldots,\boldsymbol{u}_{\text{r}},\boldsymbol{u}_{\text{o}}])$, where $\Pi(.)$ denotes the interleaving function, are encoded by the upper and lower convolutional encoders, respectively. We define the repetition ratio $\lambda \triangleq \frac{K'-K}{(q-1)K'} \in [0,\frac{1}{q}]$ as the length of $\boldsymbol{u}_r$ over $K'$. The length of $\boldsymbol{u}_{\text{o}}$ is then given by $(1-q\lambda)K'$. The codeword is a length-$N $ sequence $\boldsymbol{c}=[\boldsymbol{u}_{\text{r}},\boldsymbol{u}_{\text{o}},\boldsymbol{v}^{\text{U}},\boldsymbol{v}^{\text{L}}]=[\boldsymbol{u},\boldsymbol{v}^{\text{U}},\boldsymbol{v}^{\text{L}}]$, comprising the information sequence before repetition, as well as two length-$\frac{N-K}{2}$ parity sequences output from the upper and lower convolutional encoders, i.e., $\boldsymbol{v}^{\text{U}}$ and $\boldsymbol{v}^{\text{L}}$. Given the code rate of the mother PCC $R_0 = \frac{K'}{N'}$, where $N' = N-K+K'$ is its codeword length, the code rate of the uncoupled PCC with partial repetition is %as
\begin{align}\label{uncoupled_rate}
R_{\text{uc}}  = \frac{K'(1-(q-1)\lambda)}{N'-K'+K'(1-(q-1)\lambda)} = \frac{1-(q-1)\lambda}{\frac{1}{R_0}-(q-1)\lambda}.
\end{align}

\subsection{Construction of GSC-PCCs}
In this section, we show the construction of GSC-PCCs with partial repetition PCCs as the component codes. Note that the notations defined above are also used in this section.

The block diagram of a GSC-PCC with coupling memory $m=1$ at time instant $t$ is depicted in Fig. \ref{fig:GSC_PCC_ENC}(b). An information sequence $\boldsymbol{u}$ is divided into $L$ sequences of equal length $K$, which are denoted by $\boldsymbol{u}_t$, $t =1,\ldots,L$. We refer to $L$ as the coupling length. At time $t$, $\boldsymbol{u}_t$ is decomposed into $\boldsymbol{u}_{t,\text{r}}$ and $\boldsymbol{u}_{t,\text{o}}$, where $\boldsymbol{u}_{t,\text{r}}$ is a length-$\lambda K'$ sequence and $\boldsymbol{u}_{t,\text{o}}$ is a length-$K'(1-q\lambda)$ sequence. After sequence $\boldsymbol{u}_{t,\text{r}}$ is repeated $q$ times, the resultant length-$K'$ information sequence $[\boldsymbol{u}_{t,\text{r}},\ldots,\boldsymbol{u}_{t,\text{r}},\boldsymbol{u}_{t,\text{o}}]$ is decomposed into $m+1$ sequences of length $\frac{K'}{m+1}$, denoted by $\boldsymbol{u}^\text{U}_{t,t+j}$, $j=0,\ldots,m$. The information sequence $\boldsymbol{u}^\text{U}_{t,t+j}$ is used as a part of the input of the upper convolutional encoder at time $t+j$. At time $t$, a length-$K'$ information sequence $[\boldsymbol{u}^\text{U}_{t-m,t},\ldots,\boldsymbol{u}^\text{U}_{t,t}]$ is interleaved and encoded by the upper encoder. Meanwhile, the information sequence $[\boldsymbol{u}_{t,\text{r}},\ldots,\boldsymbol{u}_{t,\text{r}},\boldsymbol{u}_{t,\text{o}}]$ is interleaved and also decomposed into $m+1$ sequences of length $\frac{K'}{m+1}$, i.e., $\boldsymbol{u}^\text{L}_{t,t+j}$, $j=0,\ldots,m$, where $\boldsymbol{u}^\text{L}_{t,t+j}$ is used as a part of the input of the lower convolutional code encoder at time $t+j$. Note that the upper and lower interleavers are used such that the minimum distance of GSC-PCCs is not smaller than that of the underlying uncoupled codes. This is based on a similar argument for SC-PCCs in \cite[Section III]{8631116}. At time $t$, a length-$K'$ information sequence $[\boldsymbol{u}^\text{L}_{t-m,t},\ldots,\boldsymbol{u}^\text{L}_{t,t}]$ is interleaved and encoded by the lower encoder. The component codeword is a length-$N$ sequence $\boldsymbol{c}_t = [\boldsymbol{u}_t,\boldsymbol{v}^\text{U}_t,\boldsymbol{v}^\text{L}_t]$, where $\boldsymbol{v}^\text{U}_t$ and $\boldsymbol{v}^\text{L}_t$ are two length-$\frac{N-K}{2}$ parity sequences output from the upper and lower convolutional encoders at time $t$.

To initialize and terminate GSC-PCCs, we set $\boldsymbol{u}_t$ to $\boldsymbol{0}$ for $t\leq0$ and $t>L$. As a result, the code rate of the GSC-PCC with coupling memory $m$, coupling length $L$ and repetition factor $q$ is
\vspace{-1mm}
\begin{align}
R_{\text{sc}}&=\frac{KL}{NL+m(N-K)}\nonumber \\
&=\frac{K'L(1-(q-1)\lambda)}{L(N'-(q-1)\lambda)+m(N'-K')} \nonumber \\
&=\frac{L(1-(q-1)\lambda)}{L(\frac{1}{R_0}-(q-1)\lambda)+m(\frac{1}{R_0}-1)},
\end{align}
\vspace{-1mm}
\par\noindent
where $R_0=\frac{K'}{N'}$ is the code rate of the mother PCC. The code rate of the GSC-PCC approaches $R_{\text{uc}}$ in \eqref{uncoupled_rate} as $L \rightarrow \infty$.

\subsection{Comparison to Existing Codes}
There are connections between the proposed GSC-PCCs and some existing spatially coupled turbo-like codes. First, the proposed codes can be seen as a generalization of the conventional SC-PCCs \cite{8002601}. One can obtain a SC-PCC from a GSC-PCC by setting $q=1$. However, the introduction of partial repetition leads to a significant performance improvement, as shown in Section \ref{sec:DE} and Section \ref{sec:TS}. Further, the proposed codes have more flexible structure as they can reach a lower code rate by increasing the repetition factor while the lowest code rate for the original SC-PCCs is 1/3.

The proposed codes bear similarities to PIC-TCs \cite{PIC2020} whose coupled information bits are encoded (and protected) by two turbo encoders (four convolutional encoders). In fact, PIC-TCs can be seen as having a fraction of information bits repeated two times and using the copies of those information bits as the input of the turbo encoder at the succeeding time instant. In GSC-PCCs, this can happen when some of the information bits from $\boldsymbol{u}_{t,\text{r}}$ appear in $\boldsymbol{u}^\text{U}_{t,t}$ and $\boldsymbol{u}^\text{L}_{t,t}$ while their copies appear in $\boldsymbol{u}^\text{U}_{t,t+j}$ and $\boldsymbol{u}^\text{L}_{t,t+j}$, $j\in\{1,\ldots,m\}$. This means that a fraction of information bits are protected by the component PCC codewords at multiple time instants. However, the coupling of PIC-TCs is on the turbo code level (the information encoded by upper and lower encoders are the same) while GSC-PCCs are coupled on the convolutional code level (the information encoded by upper and lower encoders are different).

\section{Density Evolution Analysis on the BEC}\label{sec:DE}
In this section, we first look into the graph representation of GSC-PCCs and then derive the exact density evolution equations for the BEC. In this work, we consider a rate $R_0 = 1/3$ mother PCC built from two rate-$1/2$
recursive systematic convolutional codes.

\subsection{Graph Representation}
Turbo-like code ensembles can be represented by a compact graph \cite{8002601}, which simplifies the factor graph representation. The main idea is that each information or parity sequence in the factor graph is represented by a single variable node while a trellis constraint is represented by a factor node. An interleaver is represented by a line that crosses an edge.

\begin{figure}[t!]
	\centering
\includegraphics[width=1.7in,clip,keepaspectratio]{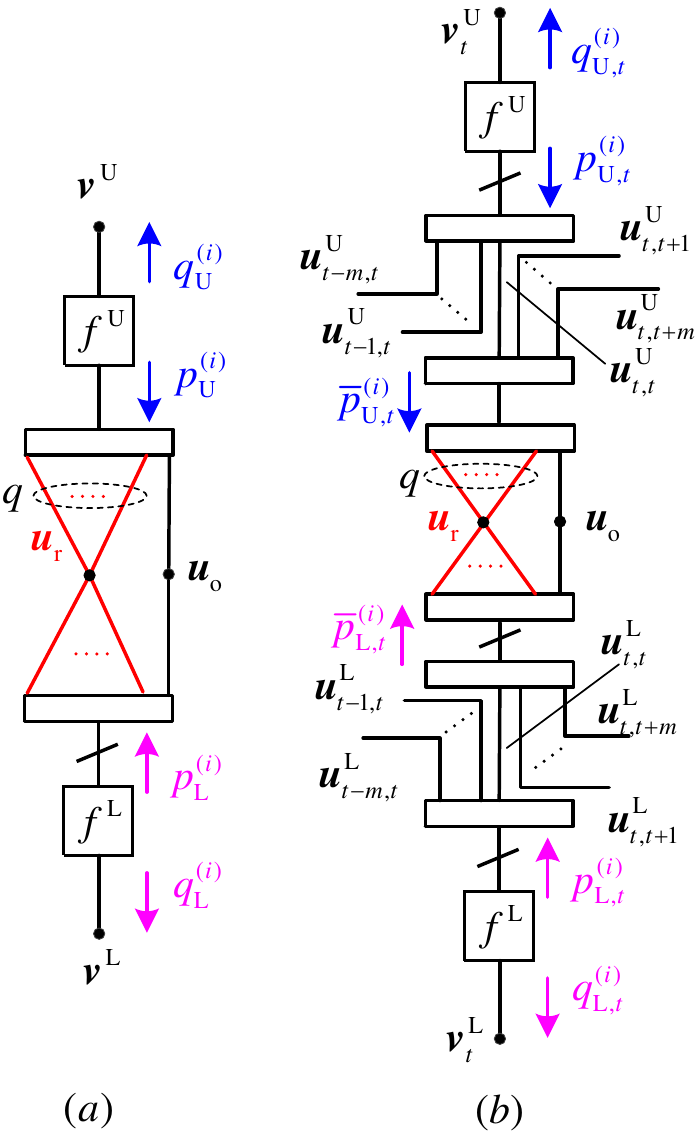}
\vspace{-3mm}
\caption{Compact graph representation of $(a)$ uncoupled ensembles, and $(b)$ GSC-PCC ensembles at time $t$.}
\label{fig:uc_graph}
\vspace{-4mm}
\end{figure}

We first look at the compact graph of an uncoupled PCC with partial repetition, which is depicted in Fig. \ref{fig:uc_graph}(a). Compared to the compact graph of the conventional PCC (see \cite[Fig. 4a]{8002601}), the difference is that in our case the information node $\boldsymbol{u}$ is represented by two nodes, $\boldsymbol{u}_{\text{r}}$ and $\boldsymbol{u}_{\text{o}}$.\footnote{With some abuse of language, we sometimes refer to a variable node
representing a sequence as the sequence itself.} Since the information sequence $\boldsymbol{u}_{\text{r}}$ is repeated $q$ times before encoded by the PCC encoder, node $\boldsymbol{u}_{\text{r}}$ connects the upper and lower factor nodes $f^\text{U}$ and $f^\text{L}$ via $q$ edges, respectively.

The compact graph representation of GSC-PCCs with coupling memory $m$ and at time $t$ is depicted in Fig. \ref{fig:uc_graph}(b). It is similar to the compact graph of SC-PCCs (see \cite[Fig. 5a]{8002601}), except that information node $\boldsymbol{u}_t$ is represented by nodes $\boldsymbol{u}_{t,\text{r}}$ and $\boldsymbol{u}_{t,\text{o}}$, where node $\boldsymbol{u}_{t,\text{r}}$ connects the upper and lower factor nodes via $q$ edges, respectively. In the next section, we derive the DE equations for the BEC based on the graphs in Fig. \ref{fig:uc_graph}.

\subsection{Density Evolution}
Let $\epsilon$ denote the channel erasure probability of the BEC. For a rate-$1/2$ convolutional code, we let $f^\text{U}_\text{s}(x,y)$ and $f^\text{U}_\text{p}(x,y)$ denote the upper Bahl–Cocke–Jelinek–Raviv (BCJR) decoder \cite{1055186} transfer functions for information and parity bits, respectively, where $x$ and $y$ correspond to the input erasure probabilities for information and parity bits, respectively. Note that on the BEC, the transfer functions can be explicitly derived by using the methods in \cite{370145,1258535}.

\subsubsection{Uncoupled Ensembles}
As shown in Fig. \ref{fig:uc_graph}(a), $p^{(i)}_{\text{U}}$ and $q^{(i)}_{\text{U}}$ represent the output erasure probability of factor node $f^{\text{U}}$ for information and parity bits, respectively, after $i$ decoding iterations. Similarly, $p^{(i)}_{\text{L}}$ and $q^{(i)}_{\text{L}}$ denote the output erasure probability of $f^{\text{L}}$ for information and parity bits, respectively.

The DE update equation for the output erasure probability of the information bits at node $f_\text{U}$ is
\vspace{-1.1mm}
\begin{align}
p^{(i)}_{\text{U}}
\hspace{-1mm} =\hspace{-1mm} f^\text{U}_{\text{s}}\left(\epsilon q\lambda \left(p^{(i-1)}_{\text{U}}\right)^{q-1}\hspace{-1mm} \left(p^{(i)}_\text{L}\right)^q\hspace{-1mm} + \epsilon\left(1-q\lambda\right)p^{(i)}_{\text{L}},\epsilon\right), \label{eq:un_de1}
%q^{(i)}_{\text{U}}
%=f^\text{U}_{\text{p}}\left(\epsilon q\lambda \left(p^{(i-1)}_{\text{U}}\right)^{q-1} \left(p^{(i)}_\text{L}\right)^q+\epsilon\left(1-q\lambda\right)p^{(i)}_{\text{L}},\epsilon\right),\label{eq:un_de2}
\end{align}
\vspace{-3.4mm}
\par\noindent
where $(1-q\lambda)$ and $q\lambda$ are the ratios of the lengths of $\boldsymbol{u}_{\text{o}}$ and $[\boldsymbol{u}_{\text{r}},\ldots,\boldsymbol{u}_{\text{r}}]$ over $K'$, respectively,
$\epsilon q\lambda(p^{(i-1)}_{\text{U}})^{q-1} (p^{(i)}_\text{L})^q$ is the average erasure probability from node $\boldsymbol{u}_{\text{r}}$ to node $f^\text{U}$, $\epsilon(1-q\lambda)p^{(i)}_{\text{L}}$ is the average erasure probability from node $\boldsymbol{u}_{\text{o}}$ to node $f^\text{U}$, and finally the average erasure probability from node $\boldsymbol{v}^\text{U}$ to node $f^\text{U}$ is $\epsilon$.

The DE update equation for the output erasure probability of the parity bits at node $f^\text{U}$ can be obtained by replacing the transfer function of $f^\text{U}_{\text{s}}$ by $f^\text{U}_{\text{p}}$. The DE update equations for node $f^\text{L}$ can be obtained by interchanging $p_{\text{U}}$ and $p_{\text{L}}$ in \eqref{eq:un_de1} and using the lower decoder transfer functions for information and parity bits, respectively.

\subsubsection{Coupled Ensembles}
Based on the compact graph in Fig. \ref{fig:uc_graph}(b), we denote by $p^{(i)}_{\text{U},t}$ and $q^{(i)}_{\text{U},t}$ the output erasure probability of $f^{\text{U}}$ for information and parity bits, respectively, at time $t$ and after $i$ decoding iterations. Similarly, $p^{(i)}_{\text{L},t}$ and $q^{(i)}_{\text{L},t}$ denote the output erasure probability of $f^{\text{L}}$ for information and parity bits, respectively. We also define the average erasure probability from $f^{\text{U}}$ and $f^{\text{L}}$ to $\boldsymbol{u}_t$ as $\bar{p}^{(i-1)}_{\text{U},t}$ and $\bar{p}^{(i-1)}_{\text{L},t}$, respectively, where
\begin{align}
\bar{p}^{(i-1)}_{\text{U},t} = \frac{1}{m+1}\sum_{j=0}^m p^{(i-1)}_{\text{U},t+j},\label{eq:sc_de_1}\\
\bar{p}^{(i-1)}_{\text{L},t} = \frac{1}{m+1}\sum_{j=0}^m p^{(i-1)}_{\text{L},t+j}.\label{eq:sc_de_2}
\end{align}

By using \eqref{eq:sc_de_1} and \eqref{eq:sc_de_2}, as well as taking into account partial repetition of information bits, we obtain the DE update for the erasure probability of the information bits at $f^{\text{U}}$ as
\begin{align}
p^{(i)}_{\text{U},t}
=&f^\text{U}_\text{s}\Bigg(\frac{\epsilon}{m+1}\sum_{k=0}^m \Big(q\lambda \left(\bar{p}^{(i-1)}_{\text{L},t-k}\right)^q \left(\bar{p}^{(i-1)}_{\text{U},t-k}\right)^{q-1} \nonumber \\
&+\left(1-q\lambda\right)\bar{p}^{(i-1)}_{\text{L},t-k}\Big),\epsilon  \Bigg) \nonumber \\
 = &f^\text{U}_\text{s}\Bigg(\frac{\epsilon}{m+1}\sum_{k=0}^m \Bigg(q\lambda \left(\frac{1}{m+1}\sum_{j=0}^m p^{(i-1)}_{\text{L},t+j-k}\right)^q \nonumber \\
 &\cdot\left(\frac{1}{m+1}\sum_{j=0}^m p^{(i-1)}_{\text{U},t+j-k}\right)^{q-1} \nonumber \\
 &+\frac{\left(1-q\lambda\right)}{m+1}\sum_{j=0}^m p^{(i-1)}_{\text{L},t+j-k}\Bigg),\epsilon  \Bigg). \label{eq:sc_de_3}
\end{align}
\vspace{-3.4mm}
\par\noindent
Due to space limitations, we omit the DE equations for the erasure probability of the parity bits at $f^{\text{U}}$ as well as the DE equations for $f^\text{L}$ as they can be easily obtained from \eqref{eq:sc_de_3}.

\subsection{Rate Compatible Random Puncturing}
To increase the code rate, we consider random puncturing of parity bits. Let $\rho\in [0,1]$ denote the fraction of surviving parity bits after puncturing.

For such a randomly punctured code sequence transmitted over the BEC with erasure probability $\epsilon$, the erasure probability of the parity sequence becomes $\epsilon_{\rho} = 1-(1-\epsilon)\rho$ \cite{7353121}. As a result, the DE equations for the punctured uncoupled and coupled ensembles can be obtained by substituting $\epsilon_{\rho} \rightarrow \epsilon$ for the average erasure probability of parity bits in \eqref{eq:un_de1} and \eqref{eq:sc_de_3}, respectively.

After puncturing, the code rates of both uncoupled and coupled ensembles (consider $L\rightarrow \infty$) become
\vspace{-1mm}
\begin{align}\label{eq:rate_punc}
R_{\text{uc}}  = R_{\text{sc}}  =\frac{1-(q-1)\lambda}{\left(\frac{1}{R_0}-1\right)\rho+1-(q-1)\lambda}.
\end{align}
\vspace{-3mm}
\par\noindent
Given a target rate, $q$ and $\lambda$, $\rho$ is uniquely determined.

\subsection{Decoding Thresholds}
We compute the decoding thresholds over the BEC by using the DE equations in the previous section. We consider 4-state, rate-$1/2$ convolutional encoders with generator polynomial $(1,5/7)$ in octal notation for both upper and lower encoders. Given a target code rate and $q$, we optimize the repetition ratio $\lambda$ (which uniquely determines $\rho$ according to \eqref{eq:rate_punc}) such that the iterative decoding thresholds are maximized. The optimized $\lambda$ and the corresponding thresholds for both uncoupled (denoted by $\epsilon_{\text{BP}}$) and coupled ensembles with coupling memory $m$ and $L \rightarrow \infty$ (denoted by $\epsilon^{(m)}_{\text{BP}}$) are reported in Table \ref{table1}. Note that the optimal $\lambda$ could be a range of values.

\begin{table}[t!]
  \centering
 \caption{Optimal $\lambda$ and Decoding thresholds of GSC-PCCs}\label{table1}
 \setlength\tabcolsep{3.7pt}
\begin{tabular}{c c c c c c}
\hline
 Rate   & $q$ & $\epsilon_{\text{BP}}$  &  $\epsilon^{(m=1)}_{\text{BP}}$   & $\epsilon^{(m=3)}_{\text{BP}}$  & $\epsilon^{(m=5)}_{\text{BP}}$  \\
 \hline
$3/4$  &  2 &0.2115  & 0.2326    & 0.2352 & 0.2352 \\
$3/4$  &  4 &0.2268  & 0.2380    & 0.2430  & 0.2443 \\
$3/4$  &  6 &0.2218  &  0.2406    &  0.2442 &  0.2457 \\
\hline
 $1/2$  & 2  & 0.4698    & 0.4907  &   0.4938 & 0.4938  \\
  $1/2$  & 4  &   0.4849   & 0.4940  &  0.4969   &  0.4978  \\
   $1/2$  & 6  &   0.4747   &  0.4952  &  0.4974   &  0.4982  \\
\hline
$1/3$  &  2 & 0.6446   & 0.6627   & 0.6647  & 0.6647 \\
$1/3$  &  4 &  0.6583   & 0.6642   &  0.6656  &  0.6660 \\
$1/3$  &  6 &    0.6512  &   0.6648  &   0.6658 &   0.6661\\
  \hline
 Rate   & $q$ & $\lambda$  &  $\lambda^{(m=1)}$   & $\lambda^{(m=3)}$  & $\lambda^{(m=5)}$  \\
 \hline
$3/4$  &  2 &$[0.287,0.313]$  & 0.5    & 0.5 & 0.5 \\
$3/4$  &  4 &0.172  & $[0.201,0.206]$    &  0.24  &  0.25 \\
$3/4$  &  6 &0.13  &   0.14   &  $[0.152,0.154]$  &   $[0.162,0.163]$ \\
\hline
 $1/2$  & 2  & $[0.184,0.213]$    & 0.44  &  0.5 & 0.5  \\
 $1/2$  & 4  &  0.147    & $[0.187,0.189]$  & 0.23  &  0.25  \\
  $1/2$  & 6  &  0.12    & 0.131   &  $[0.150,0.151]$ &  $[0.156,0.160]$  \\
\hline
$1/3$  &  2 &$[0.088,0.124]$  & $[0.37,0.39]$   & 0.5  & 0.5 \\
$1/3$  &  4 & $[0.107,0.108]$  & $[0.162,0.172]$   & $[0.216,0.229]$   & 0.25  \\
$1/3$  &  6 & $[0.104,0.105]$  &  $[0.121,0.122]$  &   $[0.138,0.146]$ & $[0.151,0.158]$  \\
  \hline
\end{tabular}
\vspace{-5mm}
\end{table}

It can be observed that the BP thresholds of GSC-PCCs improve with increasing $q$ for all the considered code rates and coupling memories. Moreover, the thresholds of GSC-PCCs surpass those of PIC-TCs \cite[Table III]{PIC2020} and SC-PCCs \cite[Table II]{8002601} for the same coupling memories and same code rates even for $q=2$. On the other hand, uncoupled PCCs with partial repetition have worse performance than coupled ensembles and their BP thresholds do not always improve with $q$. It is also worth noting that the optimal repetition ratio $\lambda$ approaches $\frac{1}{q}$ when $m$ is large.

%In the next section, we will show that when $q$ is large, the BP thresholds of GSC-PCCs approach the BEC capacity by using the property of threshold saturation.

\section{Potential Function and Threshold Saturation}\label{sec:TS}
In this section, we analytically show that threshold saturation occurs for GSC-PCCs and utilize this property to investigate the impacts of $q$ on the BP thresholds.

For uncoupled PCCs with partial repetition, we consider identical upper and lower encoders. Therefore, we can define $f_\text{s}\triangleq f^\text{U}_\text{s} = f^\text{L}_\text{s}$ and $x^{(i)}\triangleq p^{(i)}_{\text{L}} = p^{(i)}_\text{U}$. The DE equation in \eqref{eq:un_de1} can be written as a fixed point recursive equation
\vspace{-1mm}
\begin{align}\label{eq:pot1}
x^{(i)}= &f_\text{s}\bigg(q\epsilon\lambda \left(x^{(i-1)}\right)^{2q-1} \hspace{-1mm}+\epsilon(1-q\lambda)x^{(i-1)},1-(1-\epsilon)\rho \bigg) \nonumber \\
\overset{(a)}{=}&f\left( g\left(x^{(i)}\right);\epsilon \right),
\end{align}
\vspace{-3mm}
\par\noindent
where $(a)$ follows by using the definitions $f(x ; \epsilon) \triangleq f_\text{s}(\epsilon x, 1-(1-\epsilon)\rho)$ and $g(x)\triangleq q\lambda x^{2q-1}+(1-q\lambda)x$. One can easily check that the DE recursion in \eqref{eq:pot1} forms a scalar admissible system \cite[Def. 1]{6325197}.

%Note that $ f_\text{s}$ is increasing in both arguments $x,\epsilon \in (0,1]$ \cite[Lemma 1]{8002601}; $g$ satisfies $g'(x)=q\lambda((2q-1)x^{2q-2}-1)+1>1-q\lambda \geq 0$ for $x\in (0,1]$; $f(0;)$ thus the DE recursion in \eqref{eq:pot1} forms a scalar admissible system \cite[Def. 1]{6325197}.

For the above scalar admissible system, the potential function \cite[Def. 2]{6325197} is
\vspace{-1mm}
\begin{align}\label{eq:pot2}
U(x;\epsilon) =& xg(x)-G(x)-F(g(x);\epsilon) \nonumber \\
\overset{(b)}{=}& \left(q-\frac{1}{2}\right)\lambda x^{2q}+\frac{1}{2}(1-q\lambda)x^2 \nonumber \\
&- \int_0^{q\lambda x^{2q-1}+(1-q\lambda)x} f_s(\epsilon z,1-(1-\epsilon)\rho)dz,
%& = (q-\frac{1}{2})\frac{1}{q} x^{2q}- \int_0^{ x^{2q-1}} F_s(\epsilon z,1-(1-\epsilon)\rho)dz \\
%& =  (1- \frac{1}{2q})x^{2q} - \int_0^{ x^{2q-1}} F_s(\epsilon z,1-(1-\epsilon)\rho)dz
\end{align}
\vspace{-3mm}
\par\noindent
where $(b)$ follows that $F(x;\epsilon) = \int_0^x f(z;\epsilon) dz = \int_0^x f_\text{s}(\epsilon z, 1-(1-\epsilon)\rho) dz$ and $G(x) = \int_0^x g(z) dz = \frac{1}{2}\lambda x^{2q}+\frac{1}{2}(1-q\lambda)x^2$.

Similar to the uncoupled case and by letting $x^{(i)}_t\triangleq \bar{p}^{(i)}_{\text{L},t} = \bar{p}^{(i)}_{\text{U},t}$, the DE equation for GSC-PCCs \eqref{eq:sc_de_3} can be written as
\vspace{-2mm}
\begin{align}
x_t^{(i)} =& \frac{1}{1+m}\sum_{j=0}^mf_\text{s}\Bigg( \frac{\epsilon}{1+m}\sum_{k=0}^m \bigg(q\lambda \left(x_{t+j-k}^{(i-1)}\right)^{2q-1} \nonumber \\
&+(1-q\lambda)x_{t+j-k}^{(i-1)}\bigg) , 1-(1-\epsilon)\rho \Bigg) \\
\overset{(a)}=&\frac{1}{1+m}\sum_{j=0}^mf\left( \frac{1}{1+m}\sum_{k=0}^m g\left(x_{t+j-k}^{(i-1)}\right); \epsilon \right) \label{eq:pot3}.
\end{align}
Following \cite[Th. 1]{6325197}, for large enough coupling memory and any $\epsilon$ smaller than the potential threshold \cite[Def. 6]{6325197} associated with the potential function in \eqref{eq:pot2}, the only fixed point of the recursion in \eqref{eq:pot3} is $\boldsymbol{x}=\boldsymbol{0}$. Therefore, threshold saturation occurs for GSC-PCC ensembles.

\begin{table}[t!]
  \centering
 \caption{MAP thresholds of Uncoupled PCC with Partial Repetition}\label{MAP1}
  \setlength\tabcolsep{3.7pt}
\begin{tabular}{c c c c c c c c c}
\hline
 Rate  & States &   $\epsilon^{(q=2)}_{\text{MAP}}$  & $\epsilon^{(q=3)}_{\text{MAP}}$ & $\epsilon^{(q=4)}_{\text{MAP}}$ & $\epsilon^{(q=5)}_{\text{MAP}}$ & $\epsilon^{(q=6)}_{\text{MAP}}$ & $\epsilon^{(q=50)}_{\text{MAP}}$ \\  \hline
$9/10$	& 2 	 &  0.0751  & 0.0846 & 0.0888& 0.0913& 0.0928 & 0.0992\\
$9/10$	& 4 	 &  0.0882  & 0.0932 & 0.0952& 0.0963& 0.0970 & 0.0996\\
$9/10$	& 8 	&  0.0940  & 0.0966 & 0.0977& 0.0982&  0.0986& 0.0998\\
\hline
 $4/5$	& 2 	 &  0.1582  & 0.1747 & 0.1819& 0.1859& 0.1884 & 0.1987\\
 $4/5$	& 4 	 &  0.1848  & 0.1915 & 0.1941& 0.1955& 0.1964 & 0.1996\\
 $4/5$	& 8 	 &  0.1930  & 0.1962 & 0.1975& 0.1981& 0.1985 & 0.1998\\
\hline
 $3/4$ & 2     & 0.2027  & 0.2217 & 0.2298 & 0.2343& 0.2372 & 0.2486\\
 $3/4$ & 4     & 0.2352  & 0.2418 & 0.2444 & 0.2457& 0.2466 & 0.2496\\
 $3/4$ & 8     & 0.2435  & 0.2466 & 0.2477 & 0.2483& 0.2486 & 0.2498\\
\hline
 $2/3$ & 2     & 0.2811  & 0.3027 & 0.3116 & 0.3165& 0.3196 & 0.3318\\
 $2/3$ & 4     & 0.3209  & 0.3266 & 0.3288 & 0.3299& 0.3306 & 0.3330\\
 $2/3$ & 8     & 0.3282  & 0.3307 & 0.3316 & 0.3321& 0.3323 & 0.3332\\
\hline
 $1/2$ & 2   & 0.4520  & 0.4727 & 0.4809 & 0.4854  &0.4881  & 0.4987\\
 $1/2$ & 4   & 0.4938  & 0.4968 & 0.4979 & 0.4985  & 0.4988 & 0.4998\\
$1/2$ & 8   & 0.4976  & 0.4989 & 0.4993 & 0.4995  & 0.4996 & 0.4999\\
\hline
  $1/3$ & 2   & 0.6352  & 0.6493 & 0.6548 & 0.6576  &  0.6594 &0.6659\\
  $1/3$ & 4   & 0.6647  & 0.6657 & 0.6661 & 0.6662  & 0.6663  &0.6666\\
   $1/3$ & 8   & 0.6659  & 0.6663 & 0.6665 & 0.6665  & 0.6665  &0.6666\\
  \hline
\end{tabular}
\vspace{-3mm}
\end{table}

With threshold saturation, the BP thresholds of GSC-PCCs even when $q$ is very large can be easily found via computing either the potential thresholds by \cite[Def. 6]{6325197} or the MAP thresholds by using the area theorem \cite{1523540}. Consider the GSC-PCCs with identical upper and lower 2-state, 4-state and 8-state component convolutional encoders with generator polynomials $(1,1/3)$, $(1,5/7)$ and $(1,15/13)$, respectively. The MAP thresholds of uncoupled PCCs with various $q$ (denoted by $\epsilon^{(q)}_{\text{MAP}}$) for different rates are reported in Table \ref{MAP1}. To obtain a large MAP threshold, we choose $\lambda = \frac{1}{q}$ as we observe from Table \ref{table1} that this choice leads to the largest BP threshold (which converges to the MAP threshold) when $m$ is large.

Table \ref{MAP1} shows that the MAP thresholds of uncoupled PCCs with partial repetition improve as $q$ increases. The thresholds also improve as the number of states of the component convolutional codes increases. When $q$ is large, the MAP thresholds of all the ensembles approach the BEC capacity. In particular, even the MAP thresholds for the ensembles with 2-state component convolutional codes are within 0.002 to the BEC capacity when $q =50$. This suggests that the BP thresholds of GSC-PCCs may approach the BEC capacity asymptotically as $q$ tends to infinity regardless of the number of states of the component convolutional codes. In other words, one can simply increase the repetition factor $q$ to obtain a GSC-PCC with its decoding threshold very close to the BEC capacity for any given component convolutional code.

%\begin{figure}[t!]
%	\centering
%\includegraphics[width=3.43in,clip,keepaspectratio]{figures/potential_func_v1.pdf}
%\caption{Potential function of the uncoupled TC with $q=2$.}
%\label{fig:potential_fun_1}
%\end{figure}
%
%\begin{figure}[t!]
%	\centering
%\includegraphics[width=3.43in,clip,keepaspectratio]{figures/potential_func_v2.pdf}
%\caption{Potential function of the uncoupled TC with various $q$ for rate $1/2$.}
%\label{fig:potential_fun_2}
%\end{figure}

\section{Simulation Results}
\begin{figure}[t!]
	\centering
\includegraphics[width=3.4in,clip,keepaspectratio]{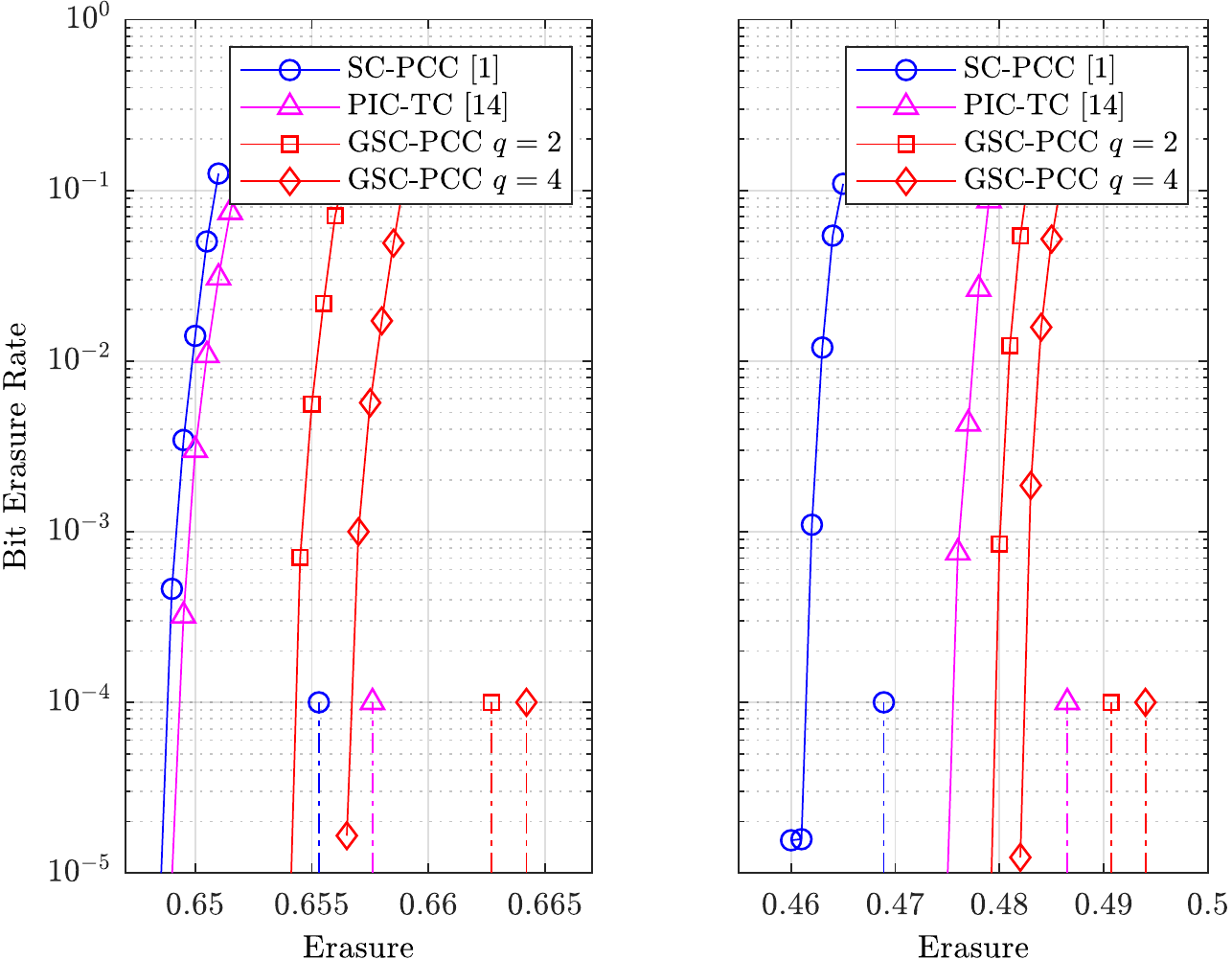}
\vspace{-2mm}
\caption{BER performance (solid lines) and density evolution thresholds (dash lines) of GSC-PCCs with rates $1/3$ and $1/2$.}
\label{fig:ber}
\vspace{-3mm}
\end{figure}

In this section, we show the finite length performance of the proposed codes on the BEC. Specifically, we consider the GSC-PCCs with identical upper and lower convolutional encoders of 4-state and with generator polynomial $(1,5/7)$. We set $K=10000$, $L=100$, $m=1$, $q\in \{2,4\}$, and $R\in\{1/3,1/2\}$. The values of $\lambda$ are chosen according to Table \ref{table1}. Moreover, we use random interleaving and randomly selected information bits for repetition and coupling (random for each transmission). The bit erasure rate (BER) and the BP thresholds for GSC-PCCs under full decoding (i.e., the decoding is performed over the whole coupled chain iteratively while standard turbo decoding is used for decoding each component codeword) are shown in Fig. \ref{fig:ber}. In the same figure, we also include the BER and decoding thresholds of SC-PCCs \cite{8002601} and PIC-TCs \cite{PIC2020} for comparison, where $R$, $K$, $L$ and $m$ are the same as those of GSC-PCCs.

We observe that for both rates, GSC-PCCs perform better than SC-PCCs and PIC-TCs and the performance gains are in agreement with the DE results. This also confirms that the optimal design of $\lambda$ is effective. Choosing $q=2$ is sufficient to allow GSC-PCCs outperform SC-PCCs and PIC-TCs while for $q=4$ the proposed codes have a noticeable performance gain over those with $q=2$. Although the BER of uncoupled PCCs is not shown in the figure, one can clearly see that the actual performance of GSC-PCCs at a BER of $10^{-5}$ is much better than the BP thresholds of uncoupled PCCs with the same $q$ (see Table \ref{table1}) or without repetition (see \cite[Table II]{8002601}). It should be noted that the BER performance of GSC-PCCs can be further improved by using a larger $q$ according to our analysis in Section \ref{sec:DE} and Section \ref{sec:TS}.

\section{Conclusions}\label{sec:conclude}
We introduced GSC-PCCs, which can be seen as a generalization of the original SC-PCCs and contain a similar structure to that of PIC-TCs. By performing a density evolution analysis on the BEC and using the potential function argument, we showed that the proposed codes exhibit threshold saturation and achieve better decoding thresholds as the repetition factor increases. Most importantly, our analysis indicates that the decoding thresholds of GSC-PCCs can approach the BEC capacity asymptotically when $q$ is large enough for any given component convolutional codes. %Our design provides an attractive way to further boost the performance of turbo-like codes.

%than existing spatially coupled turbo-like codes. Most importantly, the decoding thresholds of GSC-PCCs improve as the repetition number $q$ increases for
%
%our analysis indicates that the decoding thresholds of GSC-PCCs can be improved and made to approach the BEC capacity asymptotically by simply increasing the repetition number $q$ without changing the component convolutional codes. %

\bibliographystyle{IEEEtran}
\bibliography{MinQiu}

\begin{thebibliography}{10}
\providecommand{\url}[1]{#1}
\csname url@samestyle\endcsname
\providecommand{\newblock}{\relax}
\providecommand{\bibinfo}[2]{#2}
\providecommand{\BIBentrySTDinterwordspacing}{\spaceskip=0pt\relax}
\providecommand{\BIBentryALTinterwordstretchfactor}{4}
\providecommand{\BIBentryALTinterwordspacing}{\spaceskip=\fontdimen2\font plus
\BIBentryALTinterwordstretchfactor\fontdimen3\font minus
  \fontdimen4\font\relax}
\providecommand{\BIBforeignlanguage}[2]{{%
\expandafter\ifx\csname l@#1\endcsname\relax
\typeout{** WARNING: IEEEtran.bst: No hyphenation pattern has been}%
\typeout{** loaded for the language `#1'. Using the pattern for}%
\typeout{** the default language instead.}%
\else
\language=\csname l@#1\endcsname
\fi
#2}}
\providecommand{\BIBdecl}{\relax}
\BIBdecl

\bibitem{8002601}
S.~Moloudi, M.~Lentmaier, and A.~{Graell i Amat}, ``Spatially coupled
  turbo-like codes,'' \emph{IEEE Trans. Inf. Theory}, vol.~63, no.~10, pp.
  6199--6215, Oct. 2017.

\bibitem{397441}
C.~Berrou, A.~Glavieux, and P.~Thitimajshima, ``Near shannon limit
  error-correcting coding and decoding: Turbo-codes. 1,'' in \emph{Proc. IEEE
  Int. Conf. Commun. (ICC)}, vol.~2, May 1993, pp. 1064--1070.

\bibitem{Vucetic:2000:TCP:352869}
B.~Vucetic and J.~Yuan, \emph{Turbo Codes: Principles and Applications}.\hskip
  1em plus 0.5em minus 0.4em\relax Norwell, MA, USA: Kluwer Academic
  Publishers, 2000.

\bibitem{Gallager63low-densityparity-check}
R.~G. Gallager, ``Low-density parity-check codes,'' \emph{MIT Press}, 1963.

\bibitem{782171}
A.~{Jimenez Felstrom} and K.~S. {Zigangirov}, ``Time-varying periodic
  convolutional codes with low-density parity-check matrix,'' \emph{IEEE Trans.
  Inf. Theory}, vol.~45, no.~6, pp. 2181--2191, Sep. 1999.

\bibitem{5571910}
M.~Lentmaier, A.~Sridharan, D.~J. Costello, and K.~S. Zigangirov, ``Iterative
  decoding threshold analysis for {LDPC} convolutional codes,'' \emph{IEEE
  Trans. Inf. Theory}, vol.~56, no.~10, pp. 5274--5289, Oct. 2010.

\bibitem{5695130}
S.~Kudekar, T.~J. Richardson, and R.~L. Urbanke, ``Threshold saturation via
  spatial coupling: Why convolutional {LDPC} ensembles perform so well over the
  {BEC},'' \emph{IEEE Trans. Inf. Theory}, vol.~57, no.~2, pp. 803--834, Feb.
  2011.

\bibitem{910572}
F.~R. Kschischang, B.~J. Frey, and H.~A. Loeliger, ``Factor graphs and the
  sum-product algorithm,'' \emph{IEEE Trans. Inf. Theory}, vol.~47, no.~2, pp.
  498--519, Feb. 2001.

\bibitem{669119}
S.~{Benedetto}, D.~{Divsalar}, G.~{Montorsi}, and F.~{Pollara}, ``Serial
  concatenation of interleaved codes: performance analysis, design, and
  iterative decoding,'' \emph{IEEE Trans. Inf. Theory}, vol.~44, no.~3, pp.
  909--926, May 1998.

\bibitem{5361461}
W.~{Zhang}, M.~{Lentmaier}, K.~S. {Zigangirov}, and D.~J. {Costello}, ``Braided
  convolutional codes: A new class of turbo-like codes,'' \emph{IEEE Trans.
  Inf. Theory}, vol.~56, no.~1, pp. 316--331, Jan. 2010.

\bibitem{8631116}
S.~{Moloudi}, M.~{Lentmaier}, and A.~{Graell i Amat}, ``Spatially coupled
  turbo-like codes: A new trade-off between waterfall and error floor,''
  \emph{IEEE Trans. Commun.}, vol.~67, no.~5, pp. 3114--3123, 2019.

\bibitem{mahdavi2020effect}
\BIBentryALTinterwordspacing
M.~Mahdavi, M.~U. Farooq, L.~Liu, O.~Edfors, V.~Öwall, and M.~Lentmaier, ``The
  effect of coupling memory and block length on spatially coupled serially
  concatenated codes,'' 2020. [Online]. Available:
  \url{https://arxiv.org/abs/2006.13396}
\BIBentrySTDinterwordspacing

\bibitem{8368318}
L.~Yang, Y.~Xie, X.~Wu, J.~Yuan, X.~Cheng, and L.~Wan, ``Partially
  information-coupled turbo codes for {LTE} systems,'' \emph{IEEE Trans.
  Commun.}, vol.~66, no.~10, pp. 4381--4392, Oct. 2018.

\bibitem{PIC2020}
M.~Qiu, X.~Wu, A.~{Graell i Amat}, and J.~Yuan, ``Analysis and design of
  partially information- and partially parity-coupled turbo codes,'' \emph{IEEE
  Trans. Commun.}, vol.~69, no.~4, pp. 2107--2122, 2021.

\bibitem{8989359}
M.~Qiu, X.~Wu, and J.~Yuan, ``Density evolution analysis of partially
  information coupled turbo codes on the erasure channel,'' in \emph{Inf.
  Theory Workshop (ITW)}, Aug. 2019, pp. 1--5.

\bibitem{9174156}
X.~{Wu}, M.~{Qiu}, and J.~{Yuan}, ``Partially information coupled duo-binary
  turbo codes,'' in \emph{Proc. IEEE Int. Symp. Inf. Theory (ISIT)}, 2020, pp.
  461--466.

\bibitem{5308225}
N.~{Pillay}, H.~{Xu}, and F.~{Takawira}, ``Dual-repeat-punctured turbo codes on
  {AWGN} channels,'' in \emph{Proc. IEEE AFRICON}, 2009, pp. 1--6.

\bibitem{1055186}
L.~Bahl, J.~Cocke, F.~Jelinek, and J.~Raviv, ``Optimal decoding of linear codes
  for minimizing symbol error rate,'' \emph{IEEE Trans. Inf. Theory}, vol.~20,
  no.~2, pp. 284--287, Mar. 1974.

\bibitem{370145}
M.~R. {Best}, M.~V. {Burnashev}, Y.~{Levy}, A.~{Rabinovich}, P.~C. {Fishburn},
  A.~R. {Calderbank}, and D.~J. {Costello}, ``On a technique to calculate the
  exact performance of a convolutional code,'' \emph{IEEE Trans. Inf. Theory},
  vol.~41, no.~2, pp. 441--447, 1995.

\bibitem{1258535}
B.~M. {Kurkoski}, P.~H. {Siegel}, and J.~K. {Wolf}, ``Exact probability of
  erasure and a decoding algorithm for convolutional codes on the binary
  erasure channel,'' in \emph{Proc. IEEE Globecom}, vol.~3, Dec. 2003, pp.
  1741--1745.

\bibitem{7353121}
D.~G.~M. {Mitchell}, M.~{Lentmaier}, A.~E. {Pusane}, and D.~J. {Costello},
  ``Randomly punctured {LDPC} codes,'' \emph{IEEE J. Sel. Areas Commun.},
  vol.~34, no.~2, pp. 408--421, 2016.

\bibitem{6325197}
A.~{Yedla}, Y.~{Jian}, P.~S. {Nguyen}, and H.~D. {Pfister}, ``A simple proof of
  threshold saturation for coupled scalar recursions,'' in \emph{Proc. Int.
  Symp. Turbo Codes Iterative Inf. Process (ISTC)}, 2012, pp. 51--55.

\bibitem{1523540}
C.~{Measson}, R.~{Urbanke}, A.~{Montanari}, and T.~{Richardson}, ``Maximum a
  posteriori decoding and turbo codes for general memoryless channels,'' in
  \emph{Proc. IEEE Int. Symp. Inf. Theory (ISIT)}, Sep. 2005, pp. 1241--1245.

\end{thebibliography}

\end{document}